\documentclass[aip,reprint]{revtex4-1}
\usepackage[T1]{fontenc}
\usepackage[latin9]{inputenc}
\usepackage{babel}
\usepackage{graphicx}
\graphicspath{{imgs/}} 
\usepackage{amssymb}
\usepackage{xcolor}
\usepackage{amsmath}
\DeclareMathOperator{\sign}{sign}

\begin{document}
\title{Exploring the Interplay of Excitatory and Inhibitory Interactions in the Kuramoto Model on Circle Topologies}

\author{Albert D\'{\i}az-Guilera}
\affiliation{Departament de F\'isica de la Mat\`eria Condensada, Universitat de Barcelona, Barcelona,  Spain}
\affiliation{Universitat de Barcelona Institute of Complex Systems (UBICS), Universitat de Barcelona, Barcelona, Spain}
\author{Dimitri Marinelli}
\affiliation{Departament de F\'isica de la Mat\`eria Condensada, Universitat de Barcelona, Barcelona,  Spain}
\affiliation{Universitat de Barcelona Institute of Complex Systems (UBICS), Universitat de Barcelona, Barcelona,  Spain}
\author{Conrad J. P\'erez-Vicente}
\affiliation{Departament de F\'isica de la Mat\`eria Condensada, Universitat de Barcelona, Barcelona,  Spain}
\affiliation{Universitat de Barcelona Institute of Complex Systems (UBICS), Universitat de Barcelona, Barcelona,  Spain}

\date{\today}

\begin{abstract}

In the field of collective dynamics, the Kuramoto model serves as a benchmark for the investigation of synchronization phenomena. While mean-field approaches and complex networks have been widely studied, the simple topology of a circle is still relatively unexplored, especially in the context of excitatory and inhibitory interactions. In this work, we focus on the dynamics of the Kuramoto model on a circle with positive and negative connections paying attention to the existence of new attractors different from the synchronized state. Using analytical and computational methods, we find that even for identical oscillators, the introduction of inhibitory interactions significantly modifies the structure of the attractors of the system. Our results extend the current understanding of synchronization in simple topologies and open new avenues for the study of collective dynamics in physical systems.
\end{abstract}


\maketitle 


\section{Introduction}

The Kuramoto model (KM), first introduced by Yoshiki Kuramoto in 1975\cite{Kuramoto1975,Kuramoto1984} is a mathematical model that describes the dynamics of a system of coupled phase oscillators. It is widely used to study synchronization phenomena\cite{strogatz2000,Pikovsky2001,strogatz_sync} in a variety of physical, biological and social systems\cite{acebron2005,Rodrigues2016,fonseca2018}. The model is based on the assumption that each oscillator has a natural frequency and that the oscillators interact with each other via a diffusive nonlinear term. Much research has been done to understand under what conditions the system will synchronize. In the simplest scenario, there is a trade-off between order and disorder. Coupling favors coherence, while the distribution of frequencies in the population disrupts it. When the intensity of the interaction reaches a certain threshold, synchronization emerges spontaneously. The model has been analyzed using many different techniques over the last 40 years, and countless variants have been proposed and discussed in the literature\cite{acebron2005}.

Mean-field solutions of this model have been extensively studied under very different conditions. In these studies, topology did not play a relevant role. The basic assumption was to consider a fully connected system. However, the network topology has a large impact on the properties of the system\cite{arenas2008,dorfler2014,Rodrigues2016}.
They enrich the dynamics of synchronization phenomena. In contrast to simple lattice or fully connected network configurations, complex networks can exhibit features such as strong clustering, community structure and scale-free or small-world properties\cite{boccaletti2006}. These topological complexities have profound effects on the system's ability to achieve synchronized states. For example, hubs can act as synchronization guides and effectively control the behavior of their connected neighbors\cite{arenas2008,arenas2006b}. The focus on KM in the context of complex networks aims to show how these complicated network architectures influence collective behavior. This makes it an indispensable tool for the study of real systems where network topology is a crucial factor\cite{arenas2008,Rodrigues2016}.

In situations where all oscillators are identical, one would expect the fully synchronized state to be the only attractor of the dynamics. However, in some very specific topologies, other attractors may also appear.
For a sparsely connected network, it can be difficult to achieve full synchronization because isolated or poorly connected clusters act almost independently of each other.
This fact usually leads to new locally stable solutions with complicated basins of attraction. In contrast, a densely connected network is more likely to converge to a single synchronized state, as each oscillator is influenced by a large fraction of the others. Some authors \cite{Taylor2012,Townsend2019,Lu2020,Kassabov2022} have focused on the density of either random or regular networks and analyzed the necessary condition related to density that ensures the existence of a single attractor.

For this reason, the KM is often studied in its simplest form on a circle in order to understand the basic principles of synchronization. In this configuration, the oscillators are arranged along a circle, and each of them interacts with each of its nearest neighbors in a homogeneous way. This setup allows for analytical solutions and serves as an introduction to understanding how coupled phase oscillators can reach a phase locked state over time. The dynamical properties of these local solutions can exhibit a surprising degree of complexity. In this context, it is worth mentioning the  works that introduced the concept of winding number to classify the phase-locked states\cite{Rogge2004,Wiley2006}.
This concept is of crucial importance for the characterization of the steady-state dynamics of coupled oscillators, no matter whether they are synchronized or phase-locked states. However, determining their basins of attraction is not straightforward, as we are dealing with complex, multidimensional landscapes in which slight changes in initial conditions or network parameters can lead to completely different final states. Such complicated basin structures pose a challenge for predicting long-term behavior, but at the same time provide an opportunity to study how simple systems can lead to unexpected complex dynamics.\cite{Delabays2017,Menck2013,Zhang2021}

Most previous studies have focused on KMs with only positive (excitatory) connections between oscillators\cite{Delabays2016} covering different scenarios like unidirectional couplings\cite{Rogge2004,Seung-Yeal2012}, longer range connections\cite{Wiley2006,Townsend2019}, distribution of frequencies\cite{Ochab2010,Tilles2011}, delays\cite{denes2021}, possibility of adding or removing nodes\cite{Diaz-Guilera2008}, acyclic graphs \cite{Delabays2019}, planar graphs\cite{Delabays2017JMP,Manik2017}, etc. However, the addition of negative (inhibitory) links also leads to a more complex and diverse range of behaviors. 

In this paper, we investigate the effects of positive and negative connections on the synchronization properties of the KM. By including both excitatory and inhibitory interactions, we attempt to shed light on the interplay of these opposing forces and the resulting dynamics. We show that a circle with an arbitrary number of positive and negative connections can only be divided into two classes, despite the apparent added complexity: any circle with an even number of negative connections is equivalent to a system with all positive connections, while the circle with an odd number of negative connections is equivalent to one with a single negative link.
One of the contributions of the present work is to generalize the concept of winding number. When only positive connections are present, the winding number is an integer quantity. However, we will show later that we need to introduce a new index when the number of negative connections in a circle is odd, which takes into account the symmetry of the system and ensures that the winding number can now become half-integer.
KM has found application in many different areas, but one of the most successful is the dynamics of the brain as a first nonlinear approximation. In this field in particular, positive or negative links can play an important role, as they ensure that some attractors, such as complete synchronization in networks of identical oscillators, can no longer be achieved and other attractors appear that could possibly be associated with functional patterns in the system.\cite{Menara2022}

Let us notice that negative interactions have been also considered in the literature in a different context.
For instance, in \cite{Hong2011} the authors study a generalized KM with oscillators characterized as "conformists" and "contrarians" based on positive and negative couplings to the mean field. While our research is primarily concerned with the dynamics of positive and negative couplings, their study sheds light on related but complementary synchronization phenomena in networked systems. 

The paper is organized as follows: Section II presents the fundamentals of the model and discusses the equivalence among circles varying in the number of negative links. Section III identifies the fixed points, while their stability is examined in Sect. IV. The paper concludes by thoroughly investigating the system's attractors through extensive computer simulations and summarizing the findings.

\section{The model}
In this paper, we consider the case of identical oscillators:

$$\frac{d\theta_i}{dt} = \omega + \frac{K}{N} \sum_{j=1}^{N} a_{ij} \sin(\theta_j - \theta_i)$$
where $\theta_i$ represents the phase of the $i$-th oscillator, $\omega$ is the natural frequency, that can be taken as zero without loss of generality\cite{acebron2005}, $K$ is the coupling strength that fixes the timescale but for our purposes plays an irrelevant role and will be taken as 1, $N$ is the number of oscillators, and $a_{ij}$ are the elements of the adjacency matrix (1 if $i$ and $j$ are connected, 0 otherwise). 

\begin{figure}[h]
\begin{center}
\includegraphics[width=0.3\textwidth,angle=0]{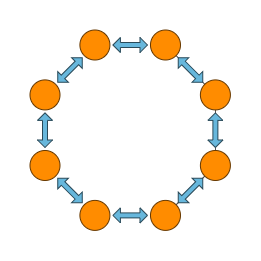}
\end{center}
\caption{The model of oscillators in a ring geometry with bidirectional couplings.}
\label{fig_circle}
\end{figure}

In the particular case of a circular geometry (see Fig. \ref{fig_circle}) each oscillator is connected to its two nearest neighbors
\begin{equation}
\dot{\theta_i}= a_{i,i+1}\sin(\theta_{i+1}-\theta_{i})+a_{i,i-1}\sin(\theta_{i-1}-\theta_{i}) \;\; \forall i=1,\ldots ,N
\label{original}
\end{equation}
where the indices of the adjacency matrix are taken mod $N$. The sign of the adjacency matrix elements can be either positive (excitatory) or negative (inhibitory).

In this particular case and taking into account the symmetry properties of the sine function the general problem with an arbitrary number of positive or negative links can be reduced to a system with only positive links or to a system with a unique negative link as we are going to show. 

Consider first that there is an even number of negative links, and 
let's introduce a new set of variables $\{ \varphi\}$ linearly related with the original set $\{ \theta\}$ as follows. 
Choose a random node in  Fig. \ref{fig_circle}, call it 0, 
and set $\varphi_0 = \theta_0$. Now let's move in a clockwise direction. Proceed with $\varphi_i = \theta_i$ until a negative link is found, 
in this case if $a_{i-1,i}<0$ make the change $\varphi_i = \theta_i + \pi$, which changes the sign of the coupling. If the following links are positive keep making changes $\varphi_j = \theta_j+\pi$ 
but when getting into a new negative link $a_{k-1,k}<0$ restore the original angles $\varphi_k = \theta_k$ and proceed in this way along the ring until another negative link is found. If the number of negative links is even, this procedure can be repeated until arriving at $i=N=0$ and $\varphi_0 = \theta_0$. The original equation has been converted into an all positive links KM for the new variables $\varphi$ such that 
\begin{equation}
    \varphi_{i}=\theta_{i}+\frac{\pi}{2}\ \left(1-\prod_{j=1}^{i}a_{j-1,j}\right)
\end{equation}
which means that depending on the number of negative links between oscillators $0$ and $i$, the change of phase will be either $0$  or $\pi$.

However, when there is an odd number of negative links this gives rise to inconsistencies,  since $i=0$ cannot be changed and hence, in this case, the problem is equivalent to a unique negative link with the same change of variable provided in the previous equation, except for the last node ($N$) that cannot be changed, and the last negative link is kept as being negative. 

In conclusion, whichever is the number of negative links we have to consider only two possible cases (either 0 or 1 negative link), and in the next sections we will discuss how this affects the possible stationary solutions or the attractors of the dynamics.

\section{Fixed points}

The fixed points correspond to the stationary solutions of the set of equations (\ref{original}) at each node. 
All the equations are formally identical, and the only difference is whether a node is in between 2 positive links or between one positive and one negative link.\footnote{Because of the equivalence described above the case of two consecutive negative links is identical to the case of two positive links}

\subsection{Local analysis}

Let us define $\Delta_{i}=\varphi_{i}-\varphi_{i-1}$.  The solution of the equation for node $i$ depends on the sign of the links around that node.

\subsubsection{Node with $++$ links}
In this case, we have
\[
0=\sin(\Delta_{i+1}) -\sin(\Delta_{i})
\]
which has two solutions

a) 
\[\Delta_{i+1}=\Delta_{i}\]

b) 
\[\Delta_{i+1}=\pi - \Delta_{i}\rightarrow \Delta_{i+1}+ \Delta_{i}=\pi\] 
which implies $\varphi_{i+1}-\varphi_{i-1}=\pi$.
Notice that solution a) involves  the three phases, but solution b) involves only the extreme phases, but not $\varphi_i$.

\subsubsection{Node with $+-$ links}
\begin{figure}[h]
\begin{center}
\includegraphics[width=0.4\textwidth,angle=0]{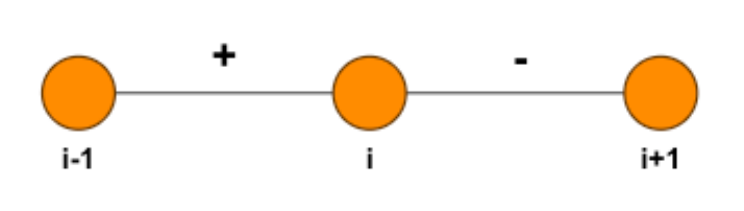}
\end{center}
\end{figure}

In this case, we have
\[
0=-\sin(\Delta_{i+1}) -\sin(\Delta_{i})
\]
which, again, has two solutions

a) 
\[\Delta_{i+1}=\Delta_{i}-\pi * \sign(\Delta_i)\]
where the sign is taken to guarantee that the phase differences are in the interval $(-\pi,\pi]$.

b) 
\[\Delta_{i+1}= - \Delta_{i} \rightarrow\Delta_{i+1}+ \Delta_{i}=0\] 
which implies
$\rightarrow$ $\varphi_{i+1}-\varphi_{i-1}=0$.
Solution a) involves again the three phases, but solution b) involves only the extreme phases, but not $\varphi_i$.

\subsection{Global analysis}
The local analysis provides information about the relation between the phases of 3 consecutive nodes. Global constraints allow us to find the possible values for the phase differences.

\subsubsection{All nodes have solution of type a)}
We have a solution such that at each node, starting by $i=1$, we assign a phase difference $\Delta_i=\varphi_i-\varphi_{i-1}=\alpha$ when the link between the nodes $i-1$ and $i$ is positive and a phase difference $\Delta_i=\varphi_i-\varphi_{i-1}=\alpha -\sign(\alpha)\pi$ when this link is negative. Which values of $\alpha$ are possible?

The sum of phase differences along the circle has to add up to an integer  multiple of $2\pi$, this is the global constraint. If we call $n$ the number of negative links, we have

\[
(N-n)\alpha +n*(\alpha-\sign(\alpha)\pi)=2\pi q
\]
where $q\in \mathbb{N}$ is defined usually as the winding number (see \cite{Wiley2006,Zhang2021}). 
We can divide by $\pi$, arrange the expression, and define the {\em lag number}, $\ell$, as
\begin{equation}    
\ell=\frac{N}{\pi}\alpha= 2q+n *\sign(\alpha)
\label{eq.ell}
\end{equation}
which we will use as specification of the state of the dynamical system (attractor) from now on. In principle, $q$ can take any integer value and the same for $\ell$. Actually, $\ell$ accounts for the symmetry of the ring. When all links are positive $\ell$ is even, but when there is one negative link then it is odd.

\subsubsection{All nodes have solution of type b)}
Solutions of type b) are such that the phases of the neighboring nodes around a given one are related, and leave the phase at the reference node free. In this case, a solution of 2 alternate subnetworks of independent phases is possible. The two subnetworks are formed by alternate nodes such that 
$$
\varphi_{i+2}-\varphi_{i}=\pi  
$$
if $a_{i,i+1}\cdot a_{i+1,i+2}=1$ and $0$ otherwise.
However, if $N$ is an odd number this decomposition is not possible and hence a global solution of type b) is not possible.
Let us assume by the moment that this solution is possible and wait for its stability in the next section.

\subsubsection{Mixing of solutions of the two types}

\begin{figure}[h!]
\begin{center}
\includegraphics[width=1.20\linewidth,angle=0.0]{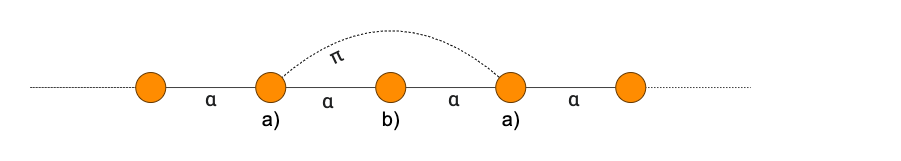}
\end{center}
\caption{Contribution of a single solution of type b) when all links are positive}
\label{fig:1btype}

\end{figure}

Let us assume first a global solution of type a) with a single solution for one node of type b) as shown in Fig. \ref{fig:1btype}.

In this case, we have $\alpha$ on the left and on the right imposed by the further nodes (labelled a)). Taking into account the type $a)$ solution at both nodes it means that the two links have the same phase difference $\alpha$, but these two phase differences have to sum $\pi$ because of type b) solution, hence we conclude that $\alpha$ has to be $\pi /2$. We have only displayed the case of two consecutive positive links, but when we have negative links the conclusion that $\alpha=\pi /2$ remains.

\begin{figure}[h]
\begin{center}
\includegraphics[width=0.95\linewidth,angle=0]{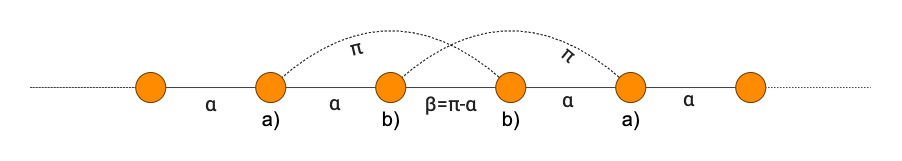}
\end{center}
\caption{Contribution of two consecutive solutions of type b) when all links are positive}
\label{fig:2btype}
\end{figure}

A nontrivial extension to the case of two consecutive $b)$ type solutions is shown in figure \ref{fig:2btype}, also for positive links.
In this case, it easily follows that between the two nodes with type b) solution the difference has to be $\beta=\pi - \alpha$. On the other hand, when one of the links between the two b) type solutions is negative it implies that the phase difference is $-\alpha$. In the next section we will discuss the stability of this type of fixed points.

Once the construction of the possible fixed points is finished, we have to look at the linear stability of the different solutions.

\section{Linear stability of the fixed points \label{Sec:Stability}}

Let's make a small perturbation in the original equations (\ref{original}) and linearize around the fixed point values of the phases: $\varphi_i = \varphi_i^* + \varepsilon_i$:
\begin{eqnarray}
\dot{\varepsilon_i}&=&
a_{i,i+1}\sin (\varphi_{i+1}^*-\varphi_{i}^* + \varepsilon_{i+1}-\varepsilon_i) + \nonumber\\
&& + a_{i,i-1}\sin (\varphi_{i-1}^*-\varphi_{i}^* + \varepsilon_{i-1}-\varepsilon_i) = \nonumber \\
&=& a_{i,i+1}\sin (\varphi_{i+1}^*-\varphi_{i}^*)\cos (\varepsilon_{i+1}-\varepsilon_i) + \nonumber \\
&& + a_{i,i+1}\cos (\varphi_{i+1}^*-\varphi_{i}^*)\sin (\varepsilon_{i+1}-\varepsilon_i)+\nonumber\\
&& +a_{i,i-1}\sin (\varphi_{i-1}^*-\varphi_{i}^*)\cos (\varepsilon_{i-1}-\varepsilon_i) +\nonumber\\
&& +a_{i,i-1}\cos (\varphi_{i-1}^*-\varphi_{i}^*)\sin (\varepsilon_{i-1}-\varepsilon_i)
\end{eqnarray}

Linearizing for small $\{\varepsilon_i\}$ the previous equation, it allows us to write it in terms of the phase differences $\{\Delta_i\}$
\begin{equation}
\dot{\varepsilon_i}=
a_{i,i+1}\cos (\Delta_{i+1}^*) (\varepsilon_{i+1}-\varepsilon_i)+
a_{i,i-1}\cos (\Delta_{i}^*) (\varepsilon_{i-1}-\varepsilon_i)
\label{perturb}
\end{equation}
\subsection{Type a) solutions}
Taking into account the value of the phase difference and how it depends on the sign we have
\[
\Delta^*_i=\alpha+\frac{1-a_{i,i+1}}{2}\cdot 
 \pi
\]
Regardless of the sign of the links between node $i$ and its neighbors, the previous equation (\ref{perturb}) reads
\[
\dot{\varepsilon_i}=
\cos (\alpha) (\varepsilon_{i-1}-2\varepsilon_i+ \varepsilon_{i+1})
\]

The set of equations can be rewritten in matrix form as
$$
\dot\varepsilon_i = \sum_{j=1}^N M_{ij} \varepsilon_j
$$

where
\[
M=\cos(\alpha)\left(\begin{array}{ccccccc}
-2 & 1 & & & & & 1 \\
1 & -2 & 1 & & & & \\
& 1 & -2 & 1 & & & \\
& & \ddots & \ddots & \ddots & & \\
& & & 1 & -2 & 1 & \\
& & & & 1 & -2 & 1 \\
1 & & & & & 1 & -2
\end{array}\right)
\]
Since the values of the elements are the same as for a totally positive cycle, we can make use of previous works that have shown that this matrix has a well known structure. 
Applying Gershgorin theorem (see appendix \ref{gershgorin}) to our matrix, it is immediate to see that the eigenvalues are bounded in the interval $[-4,0]$ with a factor $\cos \alpha$ multiplying it. It then means that when $\alpha < \pi/2$ the fixed point is stable, and unstable otherwise. This restricts the available values of $\ell$ for the steady states to $| \ell |< N/2 $. At this point it is important to recall that in our description we have used $\ell$ as the lag number, which is more appropriate than the winding number widely used in the literature. For positive links both play the same role but for an odd number of negative links the stationary solutions have a slightly different interpretation.

\subsection{1 single b) solution}

For the situation that is shown in Fig. \ref{fig:1btype} we arrived to the conclusion that the only possible value of $\alpha$ is $\pi/2$ which, according to the previous arguments, makes the eigenvalues to be zero. This holds for any number of isolated type b) solutions and hence these fixed points are not stable.

\subsection{$+$ link between two b) type solutions}
On the other hand, the possibility of two consecutive nodes, as in Fig. \ref{fig:2btype}, having a solution of type b) is still open. Let's look again in detail at the different contributions.

In this case, when $i$ corresponds to the leftmost b) type node solution $\Delta_{i+1}^*=\pi - \alpha$ and $\Delta_i^*=\alpha$ which, in the linearized equations, leads to
\begin{eqnarray}   
\dot{\varepsilon_i}&=&
\cos (\pi-\alpha) (\varepsilon_{i+1}-\varepsilon_i) 
+
\cos (\alpha) (\varepsilon_{i-1}-\varepsilon_i)\nonumber\\
&=& \cos(\alpha) (\varepsilon_{i-1}-\varepsilon_{i+1})
\end{eqnarray}

When $i$ is the rightmost b) type node then $\Delta_{i+1}^*=\alpha$ and $\Delta_i^*=\pi-\alpha$, which implies a global change in sign 
\[
\dot{\varepsilon_i}= \cos(\alpha) (-\varepsilon_{i-1}+\varepsilon_{i+1})
\]

If the link between the two b) type solutions is negative, it is compensated by the change $\pi-\alpha \rightarrow -\alpha$ and the matrix elements are identical.

Summarizing, the matrix corresponding to the linearization around these compound solution
is

\[
M=\left(\begin{array}{ccccccccc}
-2 & 1 & & & & & & & 1 \\
1 & -2 & 1 & & & & & & \\
& 1 & -2 & 1 & & & & & \\
& & \ddots & \ddots & \ddots & & \\
& & & 1 & 0 & -1 & & & \\
& & & & -1 & 0 & 1 & & \\
& & & & \ddots & \ddots & \ddots \\
& & & & & & 1 & -2 & 1 \\
1 & & & & & & & 1 & -2
\end{array}\right)
\]

Applying Gershgorin theorem, in this case the eigenvalues lie in the range $[-4,2]$, also with a factor $\cos \alpha$ meaning that there can be some positive eigenvalues making this configuration to be unstable. 

Actually, this matrix can be considered to be a Laplacian matrix in a circle that corresponds to an adjacency matrix with all elements to be +1 except a single one that is -1.
Corollary 3.7 of \cite{Pan2016} states that the number of positive eigenvalues of a matrix like the previous one is equal to the number of negative links in the Laplacian interpretation. Hence, it shows that type b) solutions are always unstable.

If more solutions of type b) are considered then we can show that for an odd number of consecutive nodes having this type of solution the only possibility is the $\alpha=\pi/2$ which makes again the solution to be metastable. For an even number of consecutive type b) solution what we have are boxes of the same type as discussed above and again apply the corollary to show that there are as many positive eigenvalues as boxes.

\section{Attractors of the dynamics}

According to our findings in the previous section, we expect only attractors such that there is a constant phase difference between neighboring nodes, either $\alpha$ if the link is positive or $\alpha - \pi$ if the link is negative, being the values of $\alpha$ restricted to a discrete set $\alpha= \pi \ell /N$ and smaller than $\pi/2$ to guarantee the stability. 
Let us take into account as well that we have performed a change of angular variables ($\{\theta_i\}\rightarrow \{\varphi_i\}$) in such a way that the problem with an arbitrary number of negative links is reduced to a simpler situation with either 0 or 1 negative links.

In the first case, only positive links, the lag number $\ell = N \alpha /\pi$ is  equal to $2q$, which means that it is even, and the restrictions of $\alpha$ limit the possible values of $\ell < N/2$.
On the other hand, for 1 negative link, $\ell = 2q + \sign (\ell) < N/2$ and it is an odd number. 
In practice, we have two families of topologies and for each family we have different sets of attractors: for an even number of negative links only attractors characterized by an even lag number are possible, while for an odd number of negative links only attractors with an odd lag number are allowed.

\begin{figure}[]
\includegraphics[width=0.98\linewidth]{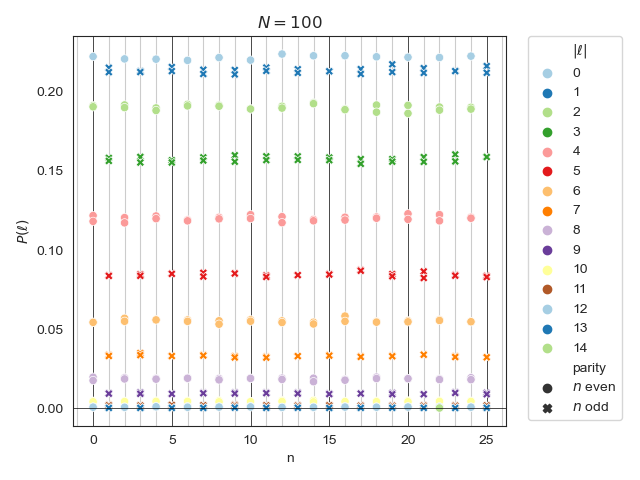}
\caption{\label{fig:equivalence}
The probability of the attractor characterized by the lag number $\ell$, $P(\ell)$, is reported on the vertical axis. For a ring of $N=100$ nodes and $n$ negative links we computed $6 \cdot 10^4$ realizations. For the sake of clarity, we neglect bins with less than 10 data points.  For $n$ even, the only accessible attractors are the ones with $\ell$ even, and their frequency is independent of $n$ (except statistical fluctuations). Conversely, $n$ odd corresponds to $\ell$ odd and $P(\ell)$ is independent of $n$ as well. 
For all configurations besides $\ell=0$ we have two almost overlapping points because, due to the symmetries of the system, $P(\ell)\sim P(-\ell)$. }
\end{figure}

We have performed simulations looking for the final stationary state of the evolution of populations of oscillators on a circle. Starting from random initial conditions, we look at the fraction of the initial states that end up in a particular final state (attractor), characterized by $\ell$. In previous works\cite{Zhang2021}, the circle with only positive links has already been analyzed in detail. Here, we generalize these results to a circle with an arbitrary number of negative links. We can corroborate by numerical simulations that an arbitrary even number of negative links is equivalent to the case where all the connections are positive. 
On the other hand, an arbitrary odd number of negative links is equivalent to the case with only one. 
\footnote{Furthermore, our simulations show that fixed points proved in Sect.\ref{Sec:Stability} to be unstable are not attractors of the dynamics.}   
In Fig.\ref{fig:equivalence} the frequency of occurrence of a final state denoted by $\ell$ shows that, for different numbers of negative links $n$, the behavior of the system differs only when the parity of $n$ changes. The simulations are performed by evolving $6 \cdot 10^4$ random initial conditions for each $(N=100,n)$ system. We solved the equations \eqref{original} with Runge-Kutta method implemented by the Julia library \texttt{DifferentialEquations.jl}\cite{rackauckas2017differentialequations} and made use of other libraries for setting up the problem\cite{NetworkDynamics.jl-2021,Graphs2021}.

Given this equivalence, we will restrict the discussion to the case $n=\{0,1\}$. To characterize the basin of attraction of the Kuramoto ring, we now focus on the statistical properties of the final states. 

\begin{figure}[h]
\begin{center}
c\includegraphics[width=0.98\linewidth,angle=0]{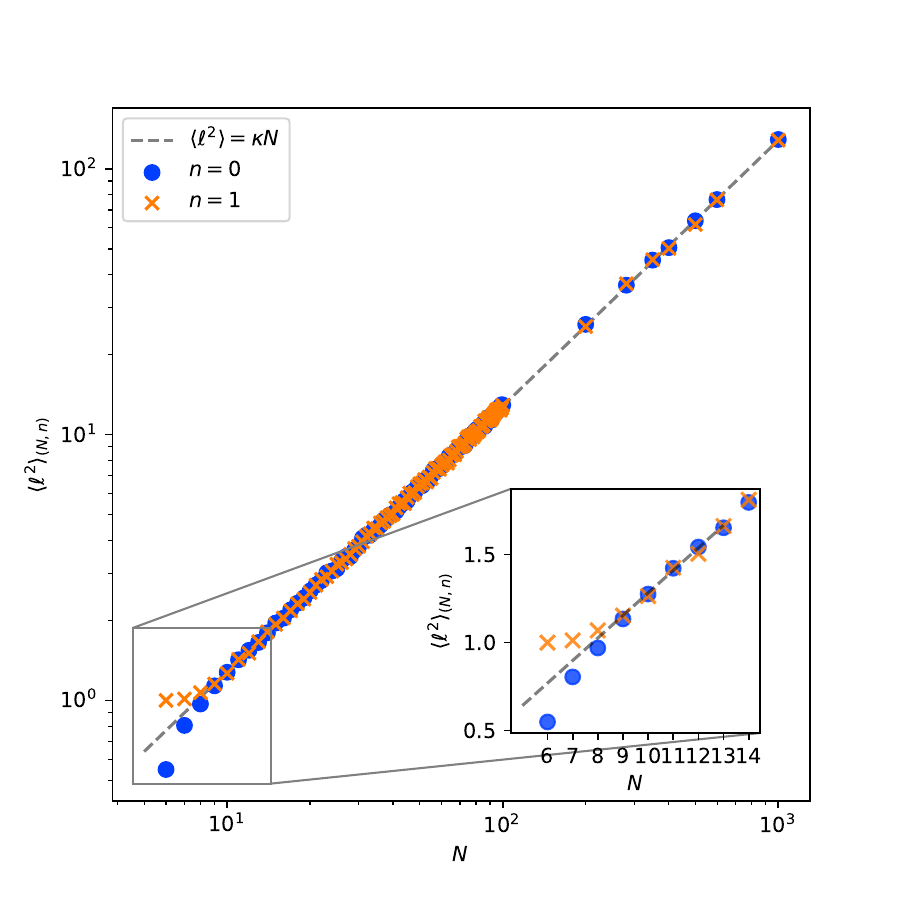}
\end{center}\vspace{-1.1cm}
\caption{\label{fig:variance_N}Second moment of the lag number $\ell$ distribution as a function of $N$. For small networks, the symmetries constraint the available values for $\left\langle \ell^{2}\right\rangle $ in case of odd or even number of negative edges.
A linear fit for $N\ge10$ shows $\kappa = 0.12807 \pm 0.0001$, with intercept compatible with zero.}
\end{figure}

\begin{figure}[h]
\begin{center}
\includegraphics[width=0.99\linewidth,angle=0]{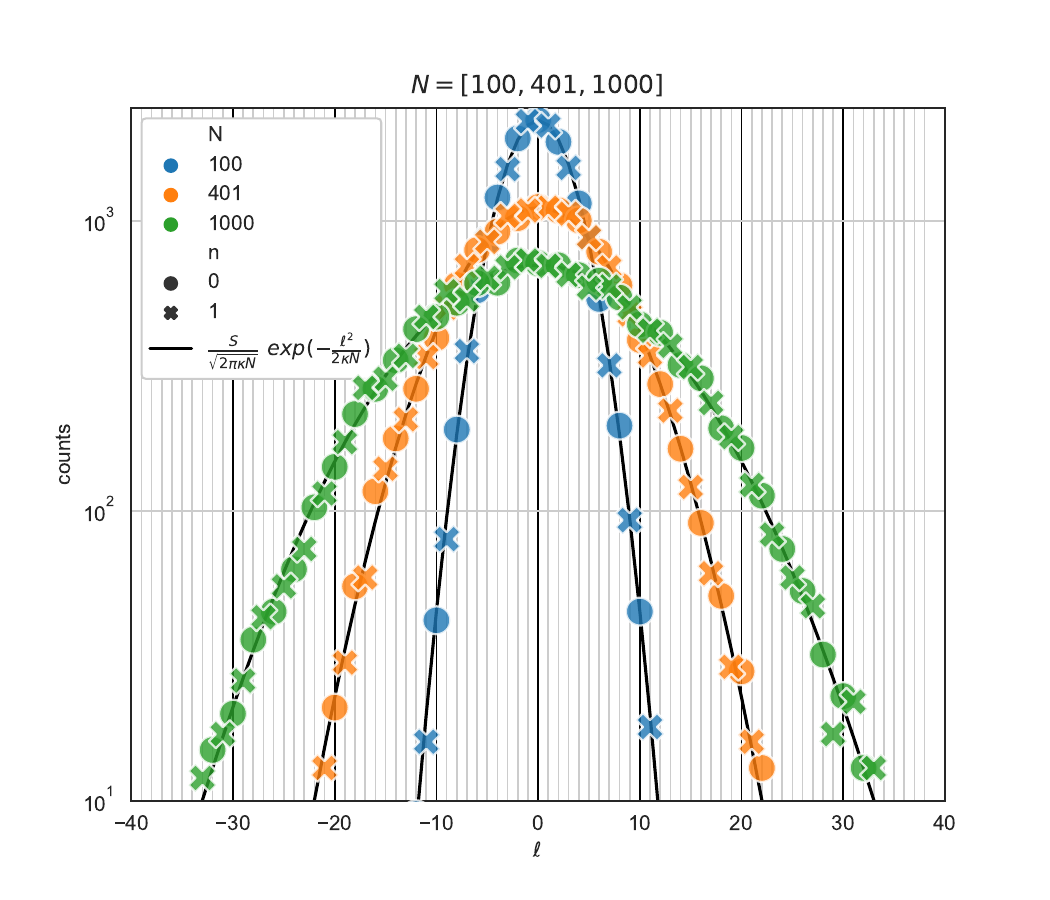} 
\end{center}

\caption{\label{fig:HistBasin}
The figure illustrates the occurrence of the basin of attraction instances characterized by the lag number $\ell$ using $\mathtt{T}/2 = 10^4$ samples with random initial conditions.  We have neglected again bins with less than 10 data points. 
The accessibility of the final states depends on the number of nodes $N$ and the parity of the number of negative edges: if $n$ even  ($n=1$) only even $\ell$ available, if $n$ odd only, odd $\ell$ only are available. The solid line represents a Gaussian distribution centered at zero with variance $\sigma^2 = \kappa N$. $\kappa$ is the only parameter whose numerical value is computed by the linear fit represented in Fig. \ref{fig:variance_N}}
\end{figure}
To study how the distribution of attractors scales with $N$, we have computed, with the method described above, the evolution of $10^4$ different initial conditions for each network configuration.
For the symmetries of the system, we expect that $E[\ell]=0$. The interesting quantity is therefore the variance $\left\langle \ell^{2}\right\rangle_{\left(N,n\right)}$. While in Fig. \ref{fig:equivalence}, one cannot see a relation between $P(\ell)$ for $n$ even and $n$ odd, one can notice in Fig. \ref{fig:variance_N}
that for $N>8$, remains the same regardless of the presence of negative links.
Moreover, the variance scales linearly with the number of nodes $N$. We can, therefore, fit a linear model and find
\begin{equation}\label{eq:linearVar}
\left\langle \ell^{2}\right\rangle_{\left(N,n\right)} = \kappa \,N    
\end{equation}
with $\kappa = 0.12807 \pm 0.0001$ and the intercept is compatible with zero. 
Looking at even moments higher than two, we can see that they are compatible with a 
Gaussian distribution centered in zero (See fig. \ref{fig:higher_moments} in appendix) 
in agreement with what was presented in \cite{Zhang2021} for $q$. 
We find, therefore, that for a fixed number of nodes, 
the attractor frequency is distributed 
along a Gaussian curve centred in the synchronized state $\ell=0$ achievable only with 
even (and zero) number of negative nodes. The curve is the same for both even and odd numbers of negative links. 
In Fig. \ref{fig:HistBasin}, we overlap the Gaussian function
\begin{gather}\label{eq:Gaussian}
   \frac{\mathtt{T}}{\sqrt{2\ \pi\ \kappa\,N}}\exp\left(-\frac{\ell^{2}}{2\;\kappa\:N}\right)
\end{gather}
 to the occurrence of the basins of attraction. $\kappa$ is the coefficient defined in eq. \eqref{eq:linearVar}, $N$ the number of nodes and $\mathtt{T}$ is the size of the sample space, namely twice the number of realizations for each configuration to take into account the two available parities. 
It should be emphasized that in contrast to a Gaussian distribution, here, $\ell$ cannot assume any value within the set of real numbers; it must be an integer,  its parity is determined by the parity of $n$ and $\ell$ is strictly bounded by $|\ell|<N/2$.

\section{Conclusions}
In this study, we have delved into the dynamics of the KM on simple circle topologies, focusing particularly on the appearance of attractors different from the synchronized state. We have expanded the traditional KM by incorporating both excitatory and inhibitory interactions among oscillators. Our analytical findings suggest that the underlying network structure profoundly influences the system's collective behavior, generating a rich tapestry of stable solutions and their corresponding basins of attraction, which are intricately dependent on the nature of the inter-oscillator interactions.

Interestingly, our work shows that even in a simple setting with identical oscillators, the introduction of inhibitory links creates a far more complex dynamical landscape than one might expect. These results challenge the conventional wisdom that assumes simplified dynamics in homogeneous systems and underscore the pivotal role of inhibitory interactions in shaping the global dynamics. Our findings open avenues for future studies aimed at understanding the consequences of these interactions in more complex networks and in systems with non-identical oscillators.

In conclusion, our study highlights the importance of considering both excitatory and inhibitory interactions when studying synchronization phenomena in networks. It invites further exploration of more complex topological structures and provides a foundation for investigating how various forms of local interactions collectively contribute to global dynamics. In the same line as the results in \cite{Zhang2021,Mihara2022,zhang2023}, we have found that the basins of attraction of the different attractors form a very intertwined structure where borders are very difficult to analyze and transitions between different attractors will be very hard to predict.

In the future, our research on the dynamics of Kuramoto oscillators on a circle with positive and negative links can be extended to explore synchronization phenomena in more complex network structures, including multiplex networks\cite{Boccaletti2014}, higher-order interactions\cite{zhang2023,Battiston2021}, and temporal networks\cite{fujiwara2011}, as suggested by previous studies. Investigating these directions holds the promise of uncovering novel synchronization patterns and behaviors, enhancing our understanding of complex systems, and enabling applications in various interdisciplinary fields, from communication networks to ecological systems, ultimately advancing our ability to model and control synchronization in real-world contexts.
\section*{Acknowledgments}
The authors acknowledge support from the Spanish grants PGC2018-094754-B-C22 and PID2021-128005NB-C22, funded by MCIN/AEI/10.13039/501100011033 and "ERDF A way of making Europe"; and from Generalitat de Catalunya (2021SGR00856). The work of D.M. has been supported by Next Generation EU through the Maria Zambrano grant from the Spanish Ministry
of Universities under the Plan de Recuperacion, Transformacion y Resiliencia.

\bibliographystyle{ieeetr}

\bibliography{general,signed_networks}

\appendix
\section{Gershgorin theorem}
\label{gershgorin}
We can make use of the Gershgorin theorem\cite{gershgorin1931,Varga2004-hn}as follows. Being $A$ a complex $n \times n$ matrix, with entries $a_{i j}$. For $i \in\{1, \ldots, n\}$ let $R_i$ be the sum of the absolute values of the non-diagonal entries in the $i$-th row:
$$
R_i=\sum_{j \neq i}\left|a_{i j}\right| .
$$
Let $D\left(a_{i i}, R_i\right) \subseteq \mathbb{C}$ be a closed disc centered at $a_{i i}$ with radius $R_i$. Such a disc is called a Gershgorin disc. Then by the Gershgorin theorem, we can say that every eigenvalue of $A$ lies within at least one of the Gershgorin discs $D\left(a_{i i}, R_i\right)$.

\section{Extended analysis}
\begin{figure*}[h]
\begin{center}
\includegraphics[width=0.98\linewidth,angle=0]{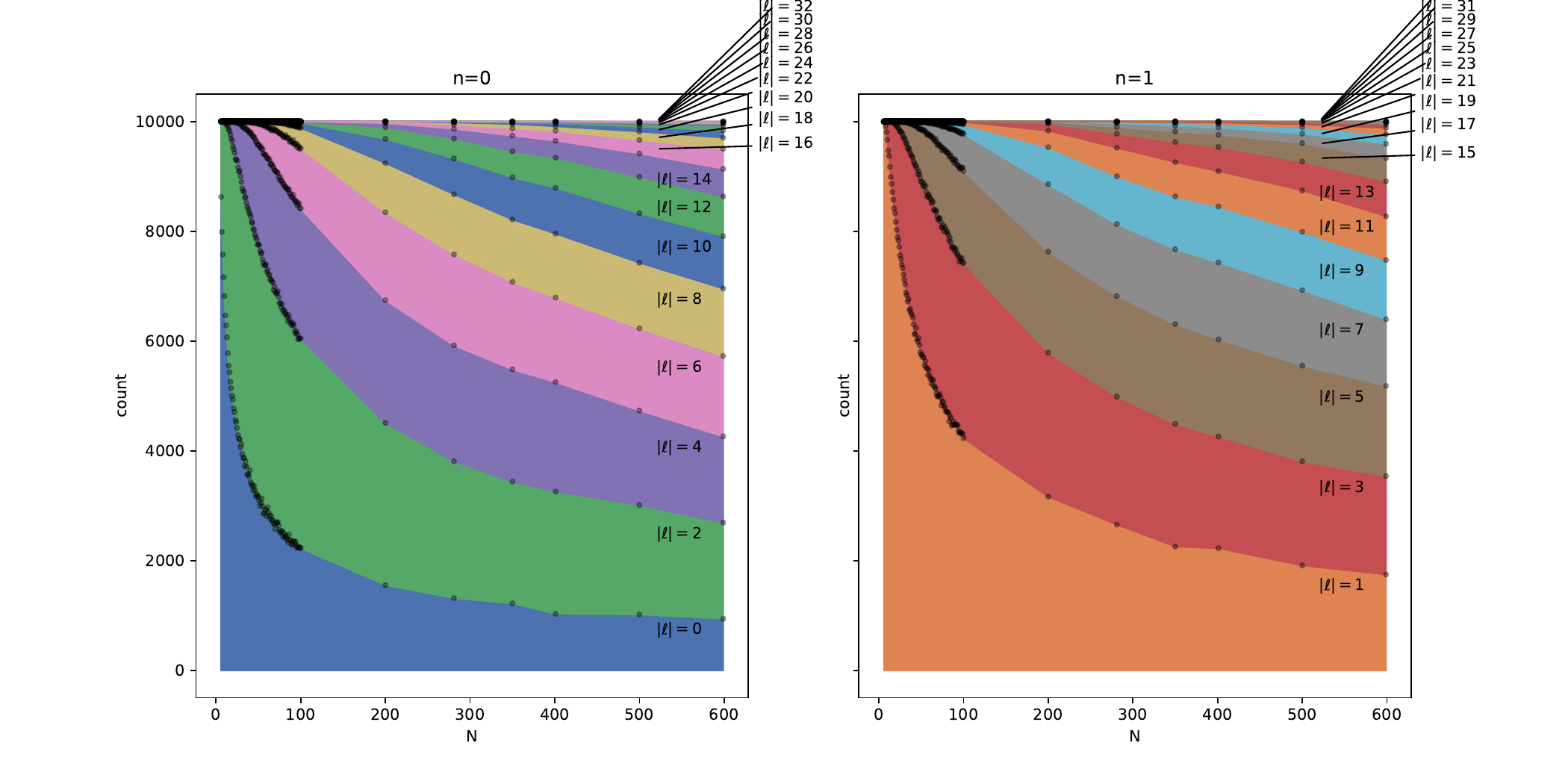}
\end{center}\vspace{-1.1cm} 
\caption{The occurrence of $10^4$ realizations of the attractors as a function of $N$ is depicted for all positive (left) and one negative (right) coupling edges. Labels corresponding to larger values of $|\ell|$ have been omitted for clarity. In this context, the area encompasses both positive and negative values of $\ell$. Consequently, the probability $P(\ell=0)$ may be lower than $P(\ell=|\ell|)$ for $\ell>0$. } 
\label{fig:dist_f_of_N}
\end{figure*}

\begin{figure}[h]
\begin{center}
\includegraphics[width=0.98\linewidth,angle=0]{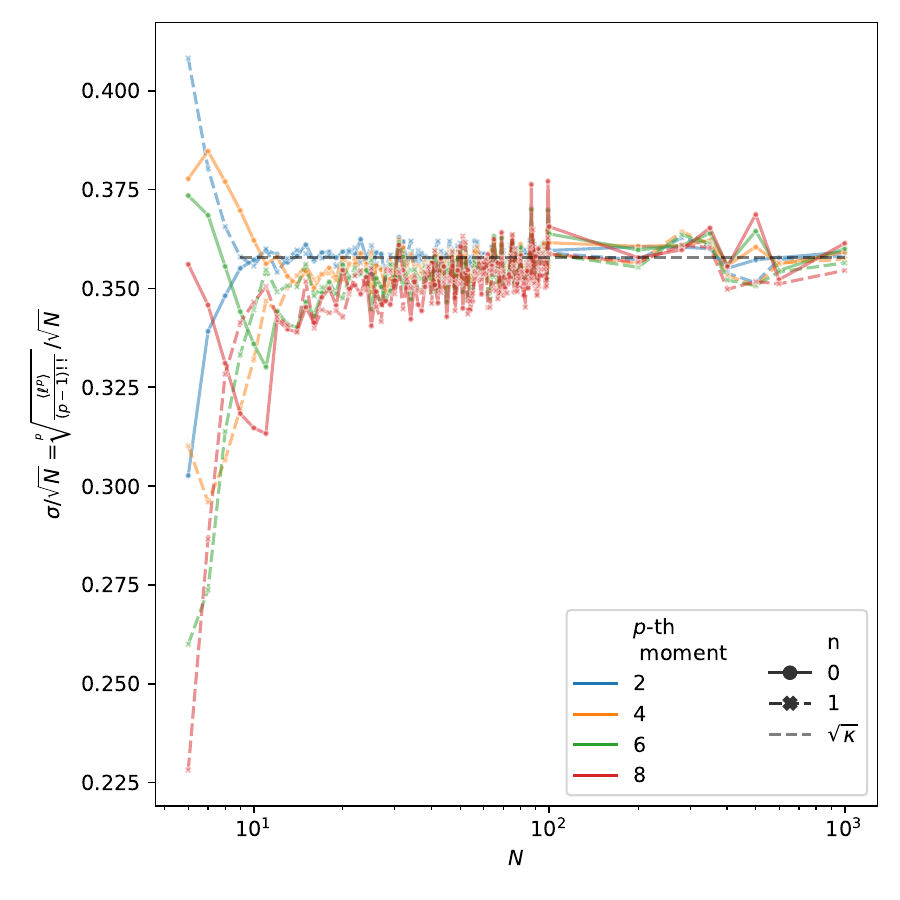}
\end{center}\vspace{-1.1cm}
\caption{\label{fig:higher_moments}
If we expect the distribution to be a Gaussian, with $\sigma$ proportional
to $\sqrt{N}$ and zero mean, we know all the higher moments.
We can compute using \eqref{eq:higher_moments} the expected $\sigma$ from higher orders:
$
\sigma=\sqrt[p]{\frac{\left\langle \ell^{p}\right\rangle }{(p-1)!!}}\qquad\text{if }p\text{ is even }
$.
}
\end{figure}

\begin{figure*}[h]
\begin{center}
\includegraphics[width=0.48\linewidth,angle=0]{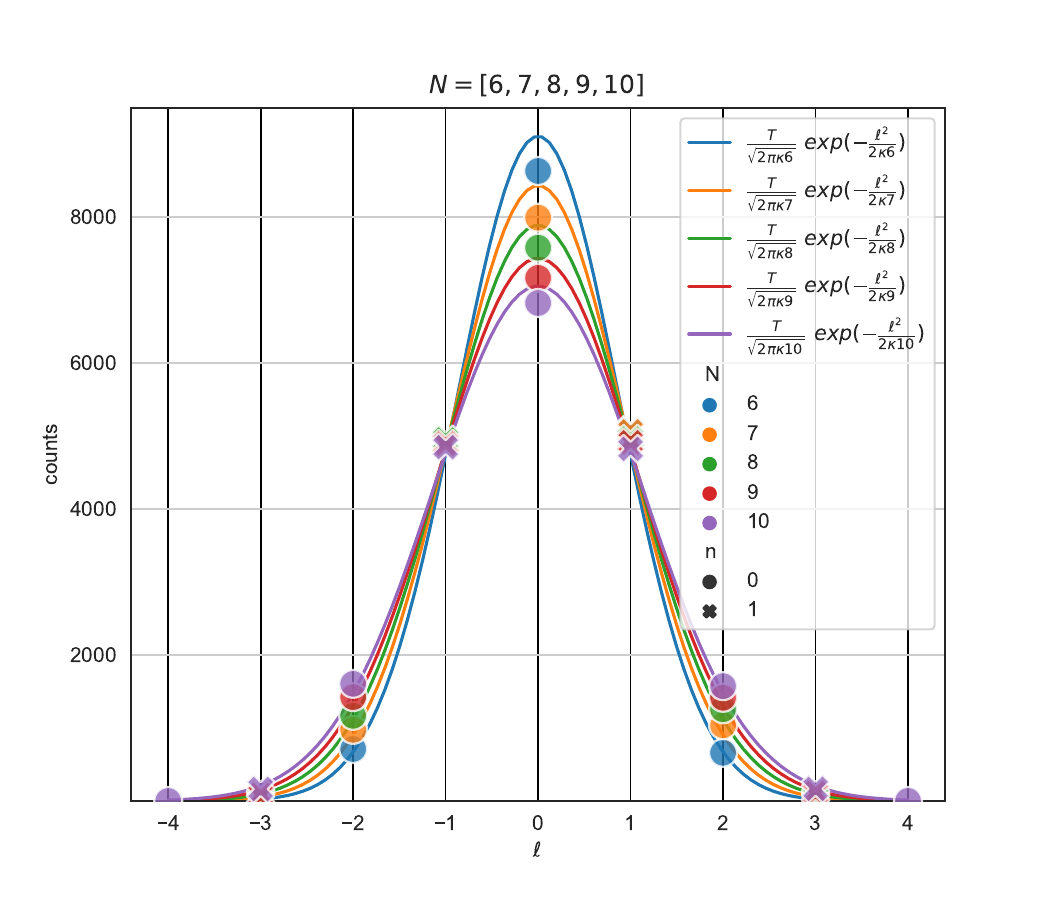}
\includegraphics[width=0.48\linewidth,angle=0]{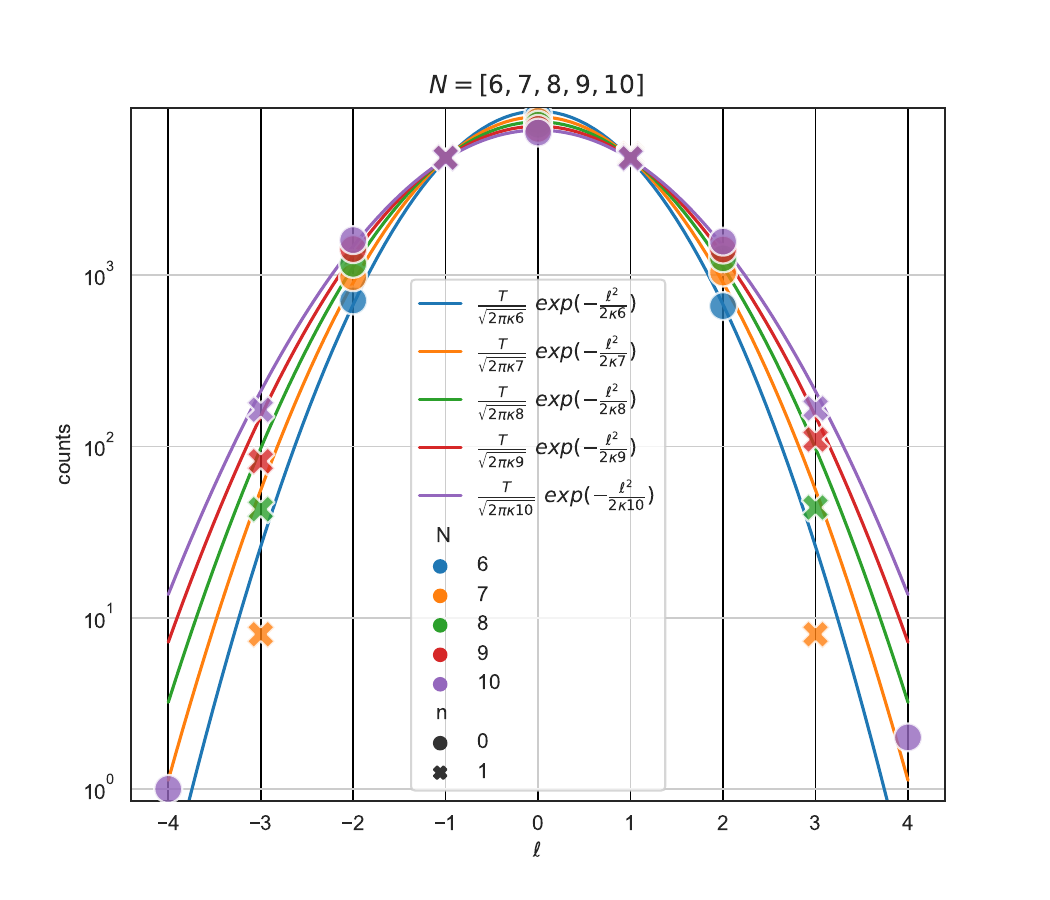}
\end{center}
\caption{\label{fig:low_N_dist}
The plot analogous to Fig. \ref{fig:HistBasin}, but for small values of $N$, illustrates that the Gaussian function inadequately approximates the distribution of the basins of attraction frequency in this case. In the left panel, the distribution is displayed on a linear scale, while in the right panel, it is represented on a logarithmic scale. In this way, we can emphasise the discrepancies in both the most and least populated states. }
\end{figure*}

We can see \eqref{eq:Gaussian} as a negative exponential of $1/N$. In this perspective in Fig. \ref{fig:dist_f_of_N}, we present how the basins of attraction distribute as a function of $N$ and for different values of $\ell$. For the realizations on the right with $n=1$, the synchronous state never occurs. 
To inspect whether, for each realization, the basins of attraction are actually distributed as Gaussian with respect to the variable $\ell$, we study higher moments. A normal-distributed random variable centred in zero, has its higher even momenta function of the second one. For $\ell$, it would read
\begin{equation}\label{eq:higher_moments}
\left\langle \ell^{p}\right\rangle _{\left(N,n\right)}=\begin{cases}
0 & \text{if }p\text{ is odd,}\\
\sigma^{p}(p-1)!! & \text{if }p\text{ is even.}
\end{cases}
\end{equation}
In Fig. \ref{fig:higher_moments}, we present a plot of the standard deviation, denoted as $\sigma$, estimated from the first eight even moments and rescaled by the system size $N$. For a normal distribution with a sufficiently large sample size, we would anticipate all data points to align along a straight line at the constant value $\sigma \sqrt{N} = \sqrt{\kappa}$, a relationship that follows from \eqref{eq:linearVar}. We can see that for larger systems, the estimation gets closer to the expected value. 
Notice that the frequency of the basins of attraction of small systems deviates from the Gaussian \eqref{eq:Gaussian}. In Fig. \eqref{fig:low_N_dist}, we can see the deviations from the expected values. The only points that overlap with the Gaussian are the $\ell=\pm 1$. For $N=6$ and $n=1$ they are the only basins of attractions available. For symmetry, we can expect that they have the same frequency and therefore $\left\langle \ell^{2}\right\rangle _{\left(6,1\right)}=1$ as we can see in Fig. \ref{fig:variance_N}.

\end{document}